\documentclass[%
 reprint,
 superscriptaddress,
 amsmath,amssymb,
 aps,
floatfix,
prfluids,
a4paper,
10pt
]{revtex4-2}

\usepackage[utf8]{inputenc}
\usepackage[english]{babel}
\usepackage[T1]{fontenc}
\usepackage{xcolor}
\usepackage{natbib}
\usepackage{booktabs}
\usepackage{placeins}
\usepackage{caption}
\usepackage{hyperref}
\hypersetup{
    colorlinks = true,
    urlcolor   = purple,
    citecolor  = purple,
    linkcolor  = purple,
}
\usepackage{orcidlink}
\usepackage[export]{adjustbox} 

\usepackage{graphicx} 
\usepackage{dcolumn} 
\usepackage{url} 
\usepackage{xurl} 

\newcommand{\micron}[0]{\mu \mathrm{m}}
\newcommand{\pd}[0]{\partial}
\newcommand{\dd}[0]{\mathrm{d}}
\newcommand{\pard}[2]{\frac{\partial #1}{\partial #2}} 

\begin{document}

\title{Under pressure: poroelastic regulation of flow in espresso brewing}

\author{Radost Waszkiewicz\orcidlink{0000-0002-0376-1708}}
\affiliation{Institute of Physics and
Astronomy, University of Potsdam, Karl-Liebknecht-Strasse 24/25, 14476 Potsdam-Golm, Germany}
\affiliation{Laboratory of Biological Physics, Institute of Physics, Polish Academy of Sciences, 02-668 Warsaw, Poland}

\author{Franciszek Myck\orcidlink{0009-0008-1364-395X}}
\affiliation{Max Planck Institute for Dynamics and Self-Organization (MPI-DS), 37077 Göttingen, Germany}
\affiliation{Faculty of Physics, University of Warsaw, L. Pasteura 5, 02-093 Warsaw, Poland}

\author{Łukasz Białas\orcidlink{0009-0008-2268-9845}}
\author{Maria Puciata-Mroczynska\orcidlink{0009-0004-2498-7679}}
\author{Michał Dzikowski\orcidlink{0000-0001-5709-7235}}
\author{Piotr Szymczak\orcidlink{0000-0001-8940-7891}}
\author{Maciej Lisicki\orcidlink{0000-0002-6976-0281}}
\email{mklis@fuw.edu.pl}
\affiliation{Faculty of Physics, University of Warsaw, L. Pasteura 5, 02-093 Warsaw, Poland}

\date{\today}

\begin{abstract}
The sensory richness of coffee is widely recognised and arises from the complex chemistry and immersion in cultural practices of coffee preparation. In contrast, the physical complexity of espresso has received less attention. The multiphase reactive flow through a dissolving, elastic porous medium remains challenging to describe. Using a controlled experimental setup based on a café-grade espresso machine, we demonstrate that the interplay between elasticity and porosity governs the long-time flow rate during espresso extraction and, consequently, the concentration of solubles in the final beverage. We introduce a minimal model that captures the resulting non-linear pressure–flow relationship and propose a methodology capable of reproducing the time-dependent behaviour of the espresso brewing process. Finally, we show that dissolution dynamics play a central role in determining the temporal evolution of flow during extraction.
\end{abstract}

\maketitle

\section{Introduction}

Espresso is a relatively recent addition to the long and diverse history of coffee preparation. Its emergence is traced back to the early 20th century \cite{Farah_2012}, in response to the growing demand for quick and concentrated coffee beverages  \cite{Morris_2010, Morris_2013}. While earlier methods, such as the cezve (or ibrik) used in the Ottoman Empire or filter-based devices popularised in European cafés, relied on extended steeping times or gentle percolation, espresso introduced a new paradigm based on pressure-driven extraction. Since its invention, espresso has evolved not only into a distinct brewing technique but also into a cultural phenomenon, particularly within the Italian caffè culture, where its preparation has acquired an almost ritualistic significance. The technical differences between all these methods lead not only to different sensory profiles, but also a different extraction efficiency and chemical composition of the final beverage \cite{Gloess_2013}. 

Among the many methods of coffee preparation, espresso remains particularly complex and often mystifying. Despite the apparent standardization of equipment and ingredients, espresso shots -- even brewed from the same coffee batch -- can vary significantly in taste and quality \cite{Hoffmann_2018}. This variability is not merely anecdotal; it reflects the sensitivity of the espresso brewing process to a range of physical parameters, including the grind size \cite{Cameron_2020,Lee_2023}, water pressure \cite{Andueza_2002}, water quality and alkalinity \cite{Navarini_2010}, temperature \cite{Andueza_2003, Salamanca_2017}, and flow rate. The extent to which this variability influences the final beverage helps explain why the underlying physics of the brewing process remains an open and active area of research.

To brew an espresso, one typically uses finely ground coffee \cite{Cameron_2020, Moroney_2015, Smrke_2024}, through which hot water is forced at high pressure, commonly around 9 bar. This pressure standard is often traced back to the force achievable in the first lever-operated espresso machines, most notably the Gaggia and Faema machines of the late 1940s and early 1950s~\cite{Morris_2010, Hoffmann_2018}. Over time, a wide array of devices and tools have been developed to aid espresso preparation, promising improved consistency, efficiency, or flavour. 
However, determining the true performance of such devices remains difficult, complicated by the subjective nature of sensory perception and the level of technical expertise and tools needed to assess brew quality.

Efforts to address these challenges have led to initiatives aimed at developing more objective and reproducible descriptors of coffee flavour and aroma. Notable among these are the World Coffee Research Sensory Lexicon \cite{Sensory_guide}, the Specialty Coffee Association’s standards for espresso brewing \cite{SCA_espresso}, and protocols for evaluating various aspects of coffee quality, as well as the widespread Coffee Taster's Flavour Wheel \cite{Spencer_2016}. These initiatives strive to bring rigour and reproducibility to a domain often clouded by marketing language and anecdotal expertise, while also promoting greater equity in the coffee value chain by enabling producers, particularly in origin countries, to better demonstrate the quality of their coffee and access premium markets through standardised, transparent assessment criteria.

\begin{figure*}
    \centering
    \includegraphics{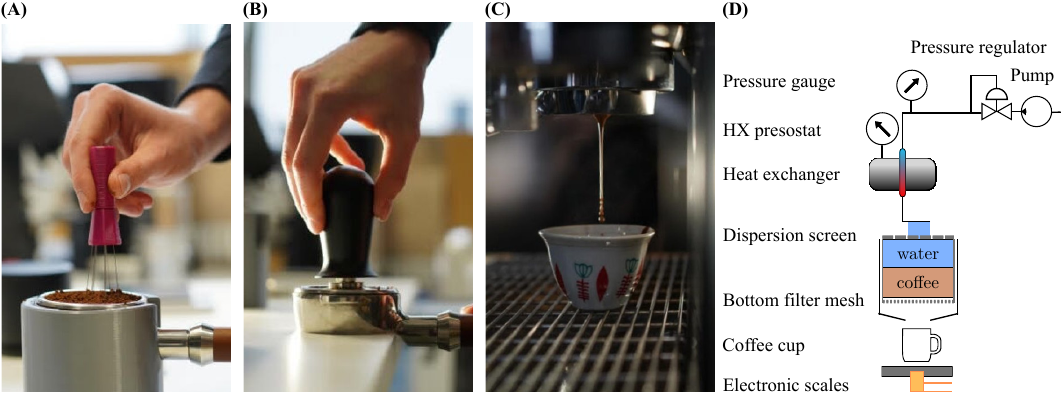}
    \caption{\textbf{The espresso brewing process.} First the weighed grounds are placed in the brewing chamber (called portafilter), then clumps are removed with a needle tool (panel \textbf{(A)}, necessary for good reproducibility). Then the grounds are tamped level with a tamper (panel \textbf{(B)}, manual tamper pictured for demonstration only) to ensure consistent preparation pressure, the espresso can then be brewed (panel \textbf{(C)}). \textbf{(D)} Simplified diagram of the heat exchange espresso machine (typical for commercial cafés). Water flows from the pump via pressure regulator, then through a heat exchanger (large amount of water in a boiler is kept at high temperature (typically 200 ℃) ensuring repeatable conditions) and into the brewing group. Inside the brewing group the water passes through a shower screen then the coffee grounds and filter basket sieve. Finally coffee ends up in the cup resting on digital scales. Data from digital pressure gauge and scales were collected simultaneously.} 
    \label{fig:preparation_and_setup}
\end{figure*}

{Espresso extraction provides a particularly rich example of reactive flow through a granular porous medium, in which processes across a wide spectrum of length and time scales are coupled. At the microscopic level, water penetrates the complex pore network of an unconsolidated and polydisperse medium that is saturated with air and residual carbon dioxide. This imbibition is accompanied by swelling of the medium and subsequent dissolution and thus--chemical erosion. At larger spatial scales, the coffee bed behaves as a confined porous matrix whose permeability might evolve locally due to swelling, compaction, erosion, or flow-induced rearrangement. These effects unfold jointly over the course of brewing, typically on the time scale of seconds, from the initial wetting and air displacement, through the release of solubles, to the long-time hydrodynamic response of the brewed puck. This confluence situates espresso brewing at the intersection of multiphase flow, poromechanics, and reactive transport, making the process challenging to characterise, model, and capture within a unified theoretical framework.} 

 To explore the multi-scale and multi-physics dynamics of the dissolving coffee bed, {involving various strategies of coarse-graining}, continuum models have been proposed for the transport of solubles and fine coffee particles~\cite{Fasano_2000,Cameron_2020,Matias_2023,Moroney_2015,Mahadevan2012,Giacomini_2020}, often combined with numerical simulations (smoothed dissipative particle dynamics, finite element method, or lattice Boltzmann models) to incorporate the effects of swelling \cite{Mo_2023}, chemical erosion \cite{Mo_2021,Mahadevan2012}, or their interplay \cite{Matias_2021,Matias_2023}. Uneven erosion can lead to the formation of channels \cite{Mahadevan2012} and uneven extraction \cite{Moroney_2019}, an undesirable process in the coffee industry. 
 
 A separate category of studies focusses on visualising and examining the structural changes accompanying the brewing, where X-ray computer tomography (CT) proves useful. By comparing the microCT images of pre- and post-extraction coffee matrices, Mo et al. \cite{Mo_2023} found that a decreasing porosity profile (from the bottom filter to the top of the puck) always develops after extraction, which can be explained their earlier model of pressure-dependent erosion \cite{Mo_2021}. Recently, Foster et al. \cite{Foster_2025} used time-resolved CT to track the infiltration front of the water into the dry coffee bed, and proposed a one-dimensional unsaturated porous medium flow model to explain the dynamics of initial wetting.   

{As argued above, brewing espresso represents a complex instance of reactive transport in a deformable porous medium. This class of problems––where flow, matrix deformation, and chemical dissolution are coupled--is well-studied in geophysics and hydrology. For instance, the feedback between dissolution and permeability evolution is a central theme in carbonate matrix acidizing~\cite{hoefner1988pore,golfier2002ability,szymczak2009wormhole}, while the coupling of flow and stress follows the foundational principles of poroelasticity established for soil consolidation~\cite{biot1941general,terzaghi1943theoretical,Coussy2003}. Unlike geological systems, however, the coffee bed is highly compliant and the timescales are rapid (seconds vs. years). In this work, we adapt these established geomechanical frameworks to the specific conditions of the espresso basket.}

Since the preparation of an espresso in a pressurised container requires a consolidated bed of finely ground coffee, an intuitive modelling choice to quantify the flow would be a Darcy-like approach with assumed constant bed permeability, used in multiple studies of coffee brewing. In earlier studies, Darcy's law coupled with a migration mechanism for fine coffee grounds has been found to explain some of the espresso flow characteristics~\cite{Fasano_2000}, but with limited experimental validation. Corrochano et al. \cite{Corrochano_2015} reported Darcy-like behaviour of coffee (with phenomenological permeability) for low pressures ($<5$ bar). In filter brew coffee, the flow is driven by hydrostatic pressure, and Darcy's law has been successfully used to predict the dynamics~\cite{Moroney_2015}, and has been employed to model a related problem of moka pot brewing \cite{Gianino_2007,King_2008,Navarini_2009}. In brewing espresso, Darcy's constitutive law was also proposed to estimate the strength of the coffee shot and extraction time~\cite{Cameron_2020}, and develop multi-physics models~\cite{Giacomini_2020}.

However, due to the complex nature of the brewing process, it is difficult to identify the sole effect of pressure on the resulting flow response of a coffee bed. {In this work, we therefore adopt a deliberately coarse-grained approach, focusing on the long-time saturated regime and isolate the contribution of poroelastic compaction to the macroscopic pressure-flow response.} We address this question {using} a combination of theoretical {analysis} and experimental validation. We propose a {coarse-grained phenomenological} model for the long-time poroelastic deformation of the coffee puck and later expand it to describe the effect of temporal evolution of the porosity upon dissolution. We perform experiments on espresso brewed in a conventional, café-grade espresso machine, in a range of parameters following the industry standards and preparation protocols. We demonstrate the {ability of our model to capture the observed pressure-flow behaviour} by examining a range of brewing pressures, where Darcy-like, linear response is seen at low brewing pressures ($<5$ bar), while the flow saturates as a function of pressure in the standard brewing regime. By performing time-resolved {solute concentration} {(i.e. total dissolved solids, TDS)} measurements on a typical espresso, we can relate the long-time brewing characteristics to the early stages of brewing, thus elucidating the details of mass transport of solubles into the cup. {The assumptions underlying this long-time description and the role of additional coupled processes during the early stages of extraction are discussed in Sec.~\ref{sec:discussion}.}

The paper is structured as follows. In Sec.~\ref{sec:experiments}, we describe the experimental details, including the characterisation of coffee samples, puck preparation procedure, details of our brewer and its calibration, measurements of coffee characteristics, and visualisation techniques. In Sec.~\ref{sec:theory}, we describe the {coarse-grained poroelastic} model. First, we focus on equilibrium, long-time pressure-flow relationship. Then, we extend our theory to include time-dependent effects of coffee bed dissolution seen at earlier times. We describe our results and compare {model} predictions to the experimental measurements in Sec.~\ref{sec:results}, followed by a more detailed discussion in Sec.~\ref{sec:discussion}. We conclude the paper in Sec.~\ref{sec:conclusions}. {Appendix \ref{sec:appendix} lists all  parameters used in the manuscript, together with their physical meaning.}

\begin{figure}[t]
    \includegraphics{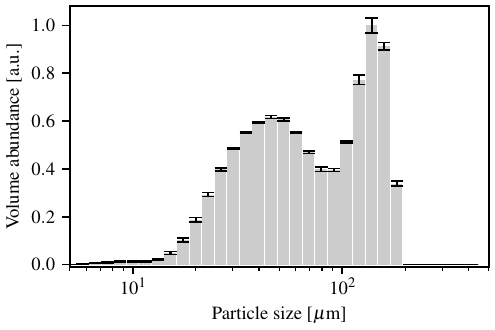} 
    \caption{\textbf{Particle abundance as a function of size.} The coffee grounds show a characteristic bimodal distribution with smaller component called ``fines'' of typical size of 50 $\mu$m and larger grounds of typical size below 200 $\mu$m.} 
    \label{fig:abundance_vs_size}
\end{figure}

\section{Experiments}\label{sec:experiments}

\subsection{Coffee Samples}
The coffee used in all measurements was a single origin specialty grade coffee from the Igarape region of Brazil, roasted by CoffeeLab Warsaw. It was a light to medium-light espresso roast, scoring 85.5 as analysed using LeBrew RoastSee C1 colour analyser.

The espresso preparation method followed the approach devised by Lance Hedrick with the help of Samo Smrke and Jonathan Gagné in \cite{Hedrick2, Hedrick}. For each brewing process recording, a fixed dose of $m_0=18.50\pm 0.05$ g was ground using Fiorenzato F64 EVO grinder at a setting of 1.9.  The grind setting was chosen based on the Specialty Coffee Association recommended brewing speed at 9 bar -- 2:1 ratio, that is the desired 37 grams of liquid in a cup to be obtained at 32 seconds, although there is a natural variability of this characteristic. The grounds were transferred into the portafilter after shaking in a Weber Workshops blind shaker for 10 seconds to ensure proper distribution, upon which a 3d-printed WDT-tool (see Fig. \ref{fig:preparation_and_setup}(A)) equipped with acupuncture needles was used to homogenise the powder distribution in the upper layer of the coffee bed. Next, the puck was tamped with a Perfex CPP-145 58mm Automatic Tamper set to a tamping force of 20 kg, ensuring repeatable tamping force. After such preparation, the coffee was brewed. This puck preparation approach was chosen to conform with the current standard within espresso brewing \cite{Hedrick}.

Ground coffee size distribution is a parameter which is crucial to flow rate and extraction \cite{Cameron_2020, Corrochano_2015}. To facilitate reproducibility of this study we analysed the size distribution of ground coffee, with a laser diffraction method. Data was obtained using Mastersizer 2000 by Malvern for three different samples of the coffee yielding the distribution visible in Fig.~\ref{fig:abundance_vs_size}. The shape follows the typical bi-modal distribution, with a maximum for `boulders' of sizes between 100-200 $\micron$ and a considerable fraction of `fines' of sizes of the order of 50 $\micron$.  

\begin{figure}
    \centering    \includegraphics{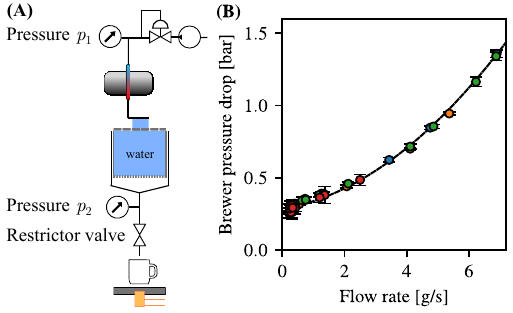}  
     \caption{
     \textbf{Brewing assembly pressure drop calibration.}
     \textbf{(A)} Espresso machine with empty basket and second pressure sensor and {restrictor valve} added below the portafilter.
     \textbf{(B)} Combined measurements from multiple series and fitted quadratic approximation.
     {Three} with a {restrictor valve} sweep and {one} with pressure regulator sweep. Brewer pressure drop depends only on the flow rate through it.
     Brewer pressure drop, $p_1-p_2$, as a function of the flow rate can be then used to obtain basket pressure from driving pressure.} 
    \label{fig:brewer_calibration}
\end{figure}

\subsection{Espresso Brewer}

For brewing, we used a two-group Sanremo Zoe Competition espresso machine, augmented with a scale under the cup (Hx711 tensometry driver), and a pressure sensor (DFRobot SEN0257), connected to a computer with a custom-built Arduino controller. A schematic illustration of the setup is drawn in Fig. \ref{fig:preparation_and_setup}(D). This allowed for accurate measurement of the mass of coffee in the cup and water pressure at the pump outlet with a frequency of 10 Hz. The parallel measurement allowed us to determine the precise moment when brewing started, which precedes the first drops falling into the cup by a couple of seconds.

{We used IMS Competizione  baskets (type B682TCH26.5E), which have an internal diameter of 58 mm, and a height of 26.5 mm. The dry coffee filled it to a height of 14 mm.}

In all experiments presented, coffee was brewed for longer than normally recommended for an espresso shot. We continued brewing each coffee for a prolonged time (typically about 120 s) to collect more precise hydrodynamic data. However, this does not affect the first 30-40 s of the brewing process required for an espresso. At longer times, we reached the steady state flow profile by pumping water through the espresso puck for 60+ seconds, whilst continuously measuring the mass of liquid that had flowed out of the machine and the driving pressure.

\subsection{Brewer Calibration}

To obtain measurements relative to the pressure just above the coffee puck rather than the pressure driving the brewing group assembly, we performed additional calibration measurements with a second pressure sensor placed below the empty brewing basket, followed by a {restrictor valve} giving variable resistance, as illustrated in Fig.~\ref{fig:brewer_calibration}(A). We performed several measurement series, varying the {restrictor valve} setting, as well as the driving pressure regulator. In this way, we found the pressure drop on the brewing group assembly as a function of the flow rate. As shown in Fig.~\ref{fig:brewer_calibration}(B), the pressure drop follows a quadratic dependence on the flow rate. {Error bars correspond to the standard deviation of the time fluctuations of the pressure signal during a given measurement, reflecting the typical experimental variability.} The non-linear hydrodynamic response of the brewer is associated with the presence of valves. The calibration curve from these measurements was used to compute the basket pressure from the raw (pump) pressure driving the brewing group, and the established relationship was subsequently used in all analyses.

\subsection{Solutes Concentration}

To characterise the dissolution process, we measured the TDS (Total Dissolved Solids) \cite{Gagne_2025} using an optical refractometer (VST-LAB coffee refractometer). To obtain time-resolved measurement, we separated an espresso shot into different containers every 5 seconds of the brewing process. The experiment was repeated three times and the results were averaged. {While the volume of an individual vial was 5 ml, thus smaller than the accumulated flow of coffee over the measurement interval, the refractometer uses only a single drop of coffee per batch for a reliable measurement.} In the case of the most concentrated brew portions, collected at early brewing times, the samples were diluted with distilled water to measurable values of TDS.

\subsection{X-ray Microtomography}

To visualise the structure of coffee pucks before and after brewing, we employ X-Ray $\mu$-CT imaging. Radiograms were collected with RX-Solutions EasyTom X-Ray tomograph, at $20\ \micron$  resolution. Image acquisition was done with the lamp set to 150kV. Scans were done just before and a couple of minutes after brewing, without removing the coffee cake from the basket.

\section{Coarse-grained poroelastic model for espresso flow}\label{sec:theory}
\subsection{Quasistatic theory}

The flow of espresso is a highly nonlinear phenomenon. However, in the long-time regime, when the dissolution process is almost complete, the flow rate through the porous bed of brewed coffee stabilises. Here we establish a relationship between the equilibrium flow rate and the applied pressure in this stable regime using a combination of the Darcy equation and a poroelastic description of the medium. {This formulation should be understood as a coarse-grained, phenomenological description of the confined coffee bed, combining established constitutive relations with a minimal representation of poroelastic deformation.}

Flow through porous media has been classically modelled with the Darcy equation that postulates a linear relationship between apparent flow $Q$ of fluid of viscosity $\mu$ through a porous medium layer of cross-section $A$ and thickness $h_0$ under a pressure difference $\Delta P$, with the proportionality constant $k$ being the permeability. Thus
\begin{equation}
    Q = k \frac{A \Delta P}{\mu h_0}.
    \label{eqn:darcy-classical}
\end{equation}
This approach can be extended to include dissolution and reactions \cite{Ladd_2021}. Deviations from the linear Darcy relationship can be modelled by allowing $k$ to be a local quantity. Since all of the phenomena under consideration depend on pressure gradients only, thus we report all pressures relative to atmospheric pressure. Further, we assume that $k$ depends only on the local porosity of the medium, $\phi$, and that its variations across the flow direction are negligible. Introducing the coordinate $z$ along the flow direction, we shall assume that all quantities are functions of $z$ only. The local formulation of Darcy flow in terms of superficial velocity $u$ is thus given by
\begin{equation}
    u(z) = {-}\frac{k(\phi(z))}{\mu}\pard{p(z)}{z},
    \label{eqn:darcy_local}
\end{equation}
where $p(z)$ and $\phi(z)$ are the depth-dependent pressure and average porosity fields. {Note that with the $z$-axis pointing upwards, $u<0$.} We postulate that at a realistic espresso brewing pressure, the coffee bed deforms, and the resulting reduction of inter-pore spaces poses additional resistance to flow. This effect of compactification of the coffee bed has been noted by coffee industry experts~\cite{Hoffmann_2018, Illy_2004}, but it has never been systematically explored or quantified. 

To construct a model that incorporates the interplay between bed permeability and deformation, we build on the ideas outlined by Hewitt et al.~\cite{Hewitt_2016}, who devised it to describe the flow of water through a saturated pack of small, soft, hydrogel spheres, driven by a pressure head. Here, we ignore the hydrostatic gradients, which are ca. $10^4$ times smaller than the driving pressure gradient. 

In the quasi-static approximation, we assume that the porosity of the coffee bed at any given time is a function of the matrix pressure $\sigma$ only, so that $\phi(z) = \phi(\sigma(z))$. Since the porous matrix is in mechanical equilibrium, the stress field $\sigma(z)$ (porous matrix pressure) inside the coffee bed is given by the Cauchy equation
\begin{equation}
    \frac{\pd}{\pd z} \left( \sigma(z) + p(z) \right) = 0.\label{eqn:mechanical_equilibrium}
\end{equation}
This formulation is also called Terzaghi's effective stress principle in the context of soil and rock poroelasticity~\cite{Guerriero_2021, Bennethum_1997}.
The mass balance of water is given by $u(z) = \textrm{const}$. {This condition follows from incompressible saturated flow at a given instant. During extraction, dissolution primarily drives a slow evolution of the effective permeability, rather than contributing significantly to the instantaneous through-flow balance, which supports a quasi-steady hydraulic description.
 In particular, the measured peak dissolved mass release rate is of order $0.1~\mathrm{g/s}$, compared with a water mass flux of order $1~\mathrm{g/s}$.}

We combine equations \eqref{eqn:darcy_local} and \eqref{eqn:mechanical_equilibrium} to arrive at a separable equation
\begin{equation}
    u = \frac{k(\phi(\sigma(z)){)} }{\mu} \frac{\pd \sigma}{\pd z},
\end{equation}
which we integrate along the coffee bed to get
\begin{equation}
   {(z - h_0)} \mu u + C=  K(\sigma), \quad K(\sigma) = \int_{0}^{\sigma}  k(\phi(\sigma'))\dd \sigma',
\end{equation}
where $h_0$ is the initial bed height and $C$ is the integration constant. From the free surface condition $\sigma(z=h_0) = 0$ we get $C = 0$. 
Note that since $k>0$, $K$ is monotonic and therefore $K^{-1}$ exists and a solution for $\sigma$ can be written as
\begin{equation}
    \sigma = K^{-1}(\mu u {(z - h_0)}).
\end{equation}
Next, imposing $\sigma(z=0) = P$ at the basket filter mesh, and recalling that $Q = {-}A u$, we find the pressure-flow relationship
\begin{equation}
     Q = K(P) \frac{A}{\mu h_0}.
     \label{eqn:solution_with_dimensions}
\end{equation}

The precise computation of $k$ from the porosity $\phi$ of a medium is a difficult task; {For the quasi-steady, saturated case studied here,} we shall assume the Carman-Kozeny relationship~\cite{Carman_1937}, typically used to approximate the permeability of granular materials, such as sand or glass beads, which yields
\begin{equation}
    k(\phi) = \frac{d_\mathrm{p}^2}{\kappa}\frac{\phi^3}{(1-\phi)^2}.\label{eqn:Darcy_Carman_Kozeny}
\end{equation}
where $d_\mathrm{p}^2$ is the effective pore diameter, and the constant $\kappa = 150$~\cite{Moroney_2019,Mccabe_1967}. 

\begin{figure}[t!]
    \includegraphics{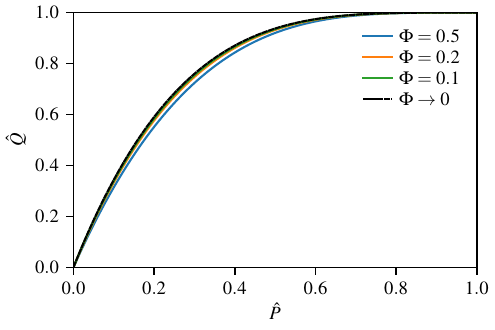}
    \caption{\textbf{Theoretical equilibrium flow rate as a function of pressure} Full theoretical expression \eqref{eqn:hat_solution_long} for normalised flow rate $\hat{Q}$ as a function of normalised pressure $\hat{P}$ at several values of control parameter $\Phi$ together with limit $\Phi \to 0$. Within the range of possible values of $\Phi$ the dependence is negligible.}
    \label{fig:phi_limit}
\end{figure}

Finally, we need to relate the matrix pressure with the porosity of the coffee bed. We assume a simplified version of the constitutive relationship used by Hewitt et al.~\cite{Hewitt_2016}, in which we keep the relationship between the matrix strain $e$ and the porosity $\phi$, given by 
\begin{equation}
 e = \frac{\Phi - \phi}{1 - \phi},\label{eqn:constitutive_relations1}
\end{equation}
where the stress-free porosity is $\Phi$, and assume a linear, Hookean relationship between the strain and the matrix stress
\begin{equation}
    \sigma = Y e, \label{eqn:constitutive_relations2}
\end{equation}
where $Y$ is the {effective} Young's modulus. In this description, $\Phi$, $Y$ and $d_p$ are the only material constants. {We stress here that the constitutive parameters above (the effective pore size and Young's modulus) should be interpreted as effective macroscopic quantities characterising the long-time, mechanically confined coffee bed, rather than fixed pore-scale material constants.} {In particular, $Y$ should be interpreted as an effective macroscopic modulus of the confined, wetted coffee bed, rather than a fixed microscopic material constant. Its value reflects the collective mechanical response of a heterogeneous, evolving granular medium subject to swelling, dissolution, and rearrangement. In this effective description, $Y$ sets the characteristic pressure scale for poroelastic compaction.}

Using these constitutive laws, we write the permeability $k(\sigma)$ as
\begin{equation}
    k(\sigma) = \frac{d_p^2}{\kappa} \frac{ (\Phi - \sigma/Y)^3}{(\Phi - 1)^2 (1-\sigma/Y)}
    \label{eqn:karman_with_dimensions}
\end{equation}
which is valid for $\sigma < \Phi Y$. When the pressure exceeds this threshold, the pores close completely.

Introducing dimensionless pressure $P^* = P / Y$,
and using Eq.~\eqref{eqn:karman_with_dimensions}, we rewrite the solution of Eq.~\eqref{eqn:solution_with_dimensions} in the dimensionless form
\begin{align}
&Q \frac{\kappa \mu h_0}{Y A d_p^2}
= (1-\Phi ) \log (1-P^*) \qquad \notag \\
&\quad + \frac{P^* \left(
        6 + 18 (\Phi -1) \Phi 
        + 3 P^* (1-3 \Phi )
        + 2 (P^*)^2
        \right)}{6 (1 -\Phi)^2}
\label{eqn:solution_dimensionless}
\end{align}

Superficially, Eq.~\eqref{eqn:solution_dimensionless} is a function of three parameters: reference flow rate $Q_\text{ref} = Y A d_p^2 / (\mu h_0)$, Young's modulus $Y$ and the porosity $\Phi$. A closer examination, however, reveals that a combination of these parameters has a negligible influence on the pressure-flow relationship. To demonstrate that, we parametrise Eq.~\eqref{eqn:solution_dimensionless} by scaling $P$ and $Q$ by their maximal allowed value, so that

\begin{equation} \label{eq:phat-qhat}
    \hat{P} = \frac{P}{Y \Phi},\qquad \hat{Q} = \frac{Q}{Q_\text{m}(\Phi)}.
\end{equation}
The maximal flow is given by
\begin{align}
    &Q_\text{m}(\Phi) = \notag \\
    &= \frac{A d_p^2 Y }{ \kappa  \mu  h_0}\frac{\Phi  (\Phi  (11 \Phi -15)+6)-6 (\Phi -1)^3 \log (1-\Phi )}{6(\Phi -1)^2}\notag \\
    &= Q_\text{ref} F(\Phi), \label{eq:F}
\end{align}
where we defined the characteristic flow rate $Q_\text{ref} = A d_p^2 Y/ \kappa  \mu  h_0$. Then, Eq.~\eqref{eqn:solution_dimensionless} takes the form
\begin{widetext}
\begin{equation}
\label{eqn:hat_solution_long}
    \hat{Q} = \frac{6 (\Phi -1)^3 \log (1-\hat{P} \Phi )
    +
    \hat{P} \Phi \left( (9\hat{P}-2\hat{P}^2-18)\Phi^2 + (18-3\hat{P})\Phi - 6 \right)
    }{
    6 (\Phi -1)^3 \log (1-\Phi ) + \Phi  ((15-11 \Phi ) \Phi -6) 
    }.
\end{equation}
\end{widetext}
As shown in Fig.~\ref{fig:phi_limit}, the dependence of $\hat{Q}$ on $\Phi$ for a given value of $\hat{P}$ is negligible. {This means that, to the leading order, the appropriately normalised long-time flow rate $\hat{Q}$ depends only on the applied pressure and not on the stress free porosity, and thus the curves of $\hat{Q}(\hat{P})$ for different values of $\Phi$, plotted in Fig.~\ref{fig:phi_limit}, vary only a little.} This result can be further understood by the examination of the series expansions of the logarithmic terms up to the fourth order. 
We are thus free to take the limit of $\Phi \to 0$ in Eq.~\eqref{eqn:hat_solution_long}, to arrive at a simplified expression
\begin{equation}    
    \hat{Q} \approx \hat{P} (4 - 6 \hat{P} + 4 \hat{P}^2 - \hat{P}^3)
    \label{eqn:teo-static-compact}
\end{equation}
without a noticeable loss of precision. Therefore, Eq.~\eqref{eqn:teo-static-compact} represents the universal shape of the long-time flow-pressure curve, independent of the value of initial bed porosity. We can use this form to determine the calibration pressure $P_\text{c}=P_\text{m}(\Phi_\text{m})$ and calibration flow $Q_\text{c}=Q_\text{m}(\Phi_\text{m})$, corresponding to the stress-free equilibrium value of porosity of the dissolved bed. {Importantly, moderate variations in microscopic constitutive properties (tortuosity, pore size, stiffness) primarily renormalise the reference scales $P_c$ and $Q_c$, while the universal shape of the nonlinear pressure--flow relation remains robust.} {In particular, variations in the effective Young's modulus do not change its functional form, reflecting the fact that the poroelastic response is governed by the ratio of applied pressure to the elastic scale set by $Y$.} The calibration values can be used to compare model results with experiments even without explicit knowledge of the value of $\Phi_\text{m}$.

\begin{figure*}[t!]
    \includegraphics{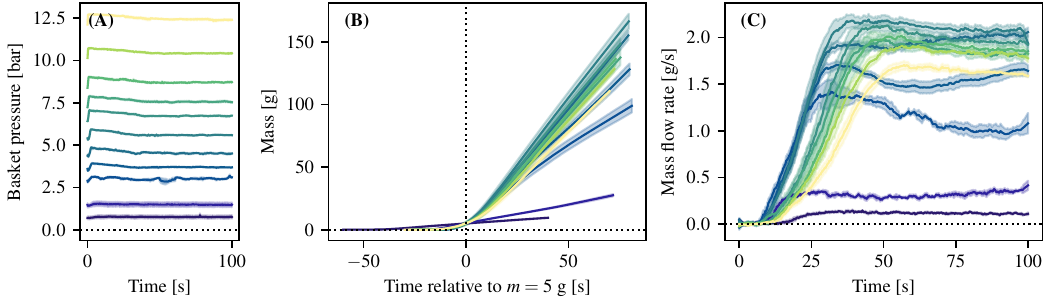}    
    \caption{\textbf{Experimental data on espresso flow at controlled pressures.} 
    \textbf{(A)} Basket pressure{, calculated from raw data using the calibration from Fig. \ref{fig:brewer_calibration},} as function of time. {Colours code the sample brewing pressure in subsequent panels}.
    \textbf{(B)} {Mass in the cup increases as a function of brewing time. Here, all plots are shifted relative to the moment when the mass in the cup equals 5 g, showing that the fastest increase corresponds to the basket pressure of ca. 5 bar. Each line is an average result of multiple experiments with shaded regions denoting standard deviation of the mean.}
    \textbf{(C)} {Mass flow rate as a function of time derived from mass measurements.
    At low pressures (dark blue) increase in pressure results in an increase in flow rate, this is contrasted with opposite trend for large pressures (light yellow).} } 
    \label{fig:time_dependent}
\end{figure*}

\subsection{Time-dependent model for early stages of brewing}

To describe the time-dependent flow rate during the early stages of dissolution (brewing), we introduce a simplified model, wherein only the stress-free matrix porosity is a function of time, $\Phi(t)$. The flow rate in Eq.~\eqref{eq:phat-qhat} can be written as
\begin{equation}
    Q = Q_\text{m}\left(\Phi(t)\right)  \hat{Q}(\hat{P}) = Q_\text{ref} F\left(\Phi(t)\right) \hat{Q}\left(\frac{P}{Y\Phi(t)}\right),
\end{equation}
{where we used the function $F$ defined in Eq.~\eqref{eq:F}.} The constants $Q_\text{ref}$ and $Y$ can be eliminated using the calibration flow and pressure to yield the final expression 
\begin{equation} \label{eq:time-dependent-q}
    Q =
    Q_c \; \frac{F(\Phi(t))}{F(\Phi_\text{m})} \; \hat{Q}\left( \frac{P \Phi_\text{m}}{P_c \Phi(t)} \right).
\end{equation}

We now assume $\Phi(t)$ to be the ratio of dissolved solids $m_\text{d}$ to the initial dose of ground coffee $m_0$: $\Phi(t) = m_\text{d}(t) / m_{0}$. The dissolved mass $m_\text{d}(t)$ can be estimated from experiments measuring the concentration of solutes as a function of time, combined with flow rate experiment performed at the same experimental conditions. 

\section{Results}\label{sec:results}

Using our experimental setup (cf. Fig. \ref{fig:preparation_and_setup}) we conducted measurements of espresso coffees brewed at different water pressures ranging from ca. 1 bar and up to 12 bar. {We present results in terms of basket pressure, which is calculated from the measured pump pressure as described in Sec. \ref{sec:experiments}}. The evolution of mass in the cup and the mass flow rate is presented in Fig.~\ref{fig:time_dependent}, where pressures are coded by colours in panel (A). In panel~\ref{fig:time_dependent}(B), we shifted the brewing time so that the point corresponding to the mass of 5 g in the cup is taken as $t=0$. We see non-monotonic dynamics, where the brews at lowest pressures exhibit very slow flow, but upon increasing the pressure above a certain threshold, the flow rate decreases again. {We then derived the mass flow rate from the mass measurements. The {measurement} was taken at 10 Hz and then smoothed with a Savitzky-Golay filter~\cite{Savitzky1964} with a window of 3 s to de-noise the derivative. Measurements were repeated several times and averaged. The cross-run standard deviation of the mean is used as the uncertainty measure in panels (B) and (C)}. From the mass flow rate evolution curves in panel~\ref{fig:time_dependent}(C), we can delineate three regimes of brewing an espresso. Firstly, the water front has to penetrate the coffee bed and expel the air entrapped within the pores. At the same time, the wetting front is accompanied by gradual swelling of the coffee bed. This process takes about $5-10$ s, during which percolation is achieved and the first drops of coffee fall into the cup. Next, we see an almost linear increase in the flow rate, which extracts material from the now wetted coffee bed. {This is in line with time-resolved X-ray observations of the wetting front~\cite{Foster_2025}, which confirm that the coffee becomes effectively saturated with water on a similar time scale under typical espresso conditions.}. At times which are past the typical brewing time of an espresso shot, we see that the flow rate saturates, or even slightly decreases towards a stable (equilibrium) value. 

\begin{figure}
    \includegraphics{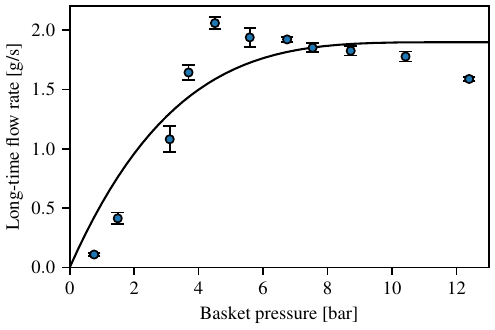}
    \caption{\textbf{Experimental long-time flow rate as a function of pressure}. Symbols indicate flow rate values {around $100$~s} after brewer activation, averaged over multiple experimental runs (error bars show the standard error of the mean), together with the fitted model prediction of Eq.~\eqref{eqn:teo-static-compact} that assumes pressure-induced compactification of the coffee bed and predicts flow rate saturation at higher pressures, while reproducing the linear, Darcy behaviour for low pressure gradients.}
    \label{fig:pressure_dependent}
\end{figure}

We first focus on this stable flow. The non-linear dependence of the long-time flow rate on the applied basket pressure can be explained using the quasi-static poroelastic model introduced in Sec.~\ref{sec:theory}.A. Assuming that the fully dissolved coffee bed remains elastic, we can compare the theoretical flow-pressure relationship, Eq.~\eqref{eq:phat-qhat} with experimental data. To this end, we brewed a total of 60 coffees at 11 different values of the basket pressure. Each coffee was brewed for at least 120 s, with the puck preparation protocol described in Sec.~\ref{sec:experiments}. The equilibrium flow rate was taken to be the average value from the last 10 s of each experimental run. The results are presented in Fig.~\ref{fig:pressure_dependent}, together with a fitted theoretical curve. The fitting parameters are the calibration flow $Q_c=1.90\pm0.15$ g/s and the calibration pressure $P_c=12\pm3$ bar. At low basket pressures, the model reproduces Darcy-like flow reported in literature before \cite{Mo_2022} for comparable pressures. However, for higher pressures the flow {saturates}, which is consistent with earlier empirical reports of coffee professionals. We note that a similar behaviour was also observed in the percolation models of Fasano {\it et al.}~\cite{Fasano_2000}, where they introduced a non-linear flow-dependent threshold for dissolution.

\begin{figure*}
    \centering
     \includegraphics{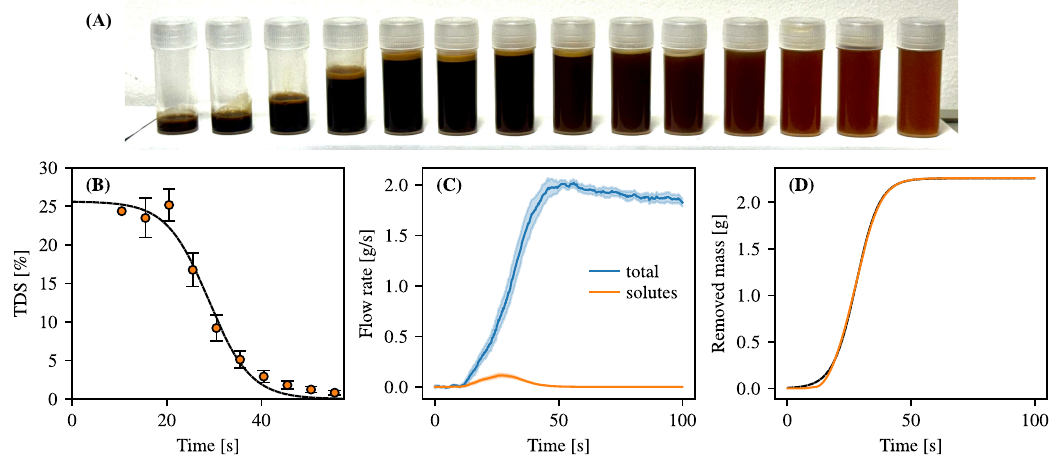}
    \caption{
    \textbf{Determination of dissolution kinetics}
    \textbf{(A)} A single espresso split into 14 fractions of 5 seconds each. Flow rate during first 20 seconds is much smaller than in later intervals.
    \textbf{(B)} Total dissolved solids (TDS) derived from refractometric measurements together with fitted approximation of Eq.~\eqref{eq:tds}.
    \textbf{(C)} Total coffee flow rate (blue) and solutes flow rate (orange), {i.e. the total flow rate multiplied by the approximate TDS at a given time}. Solutes flow is concentrated in the time of rapidly increasing total flow rate.
    \textbf{(D)} Total {mass} of removed solids {$m_\text{d}(t)$} as a function of time - computed from the solutes flux (orange) together with a fitted smooth approximation (gray).}
    \label{fig:tds}
\end{figure*}

\begin{figure*}[t]
\includegraphics{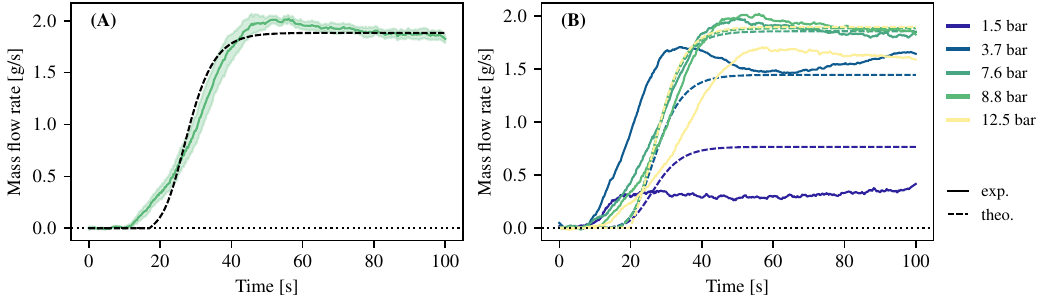}    
    \caption{\textbf{Dissolution induced changes in porosity.} \textbf{(A)} Comparison of time dependent flow profile derived from the exponential dissolution model for the most typical (9 bar) brewing pressure. \textbf{(B)} Comparison of several time dependent trajectories at different pressures. {Dashed lines symbolise theoretical flow profile estimates, while solid lines are the experimental results. The pressure colour coding is kept the same as in Fig.~\ref{fig:time_dependent}.}}
    \label{fig:porosity_model}
\end{figure*}

We now turn to the time-dependent effects in the flow of espresso. The early stages of brewing result from an interplay between the wetting and subsequent swelling of coffee, which results in a delayed flow response to applied pressure, and the dissolution kinetics. To visualise this competition, we brewed an espresso and divided the brew into 5 s intervals, illustrated in Fig.~\ref{fig:tds}(A). The first tube ($0-5$ s) was empty, and the flow rate increased systematically, beyond the capacity of a single tube {(5 ml)} in a 5 s window. At the same time, the brightening colour of the brew at longer times indicates a decrease in the amount of dissolved material. To quantify the dissolution kinetics, we measured the TDS of subsequent samples, as shown in Fig.~\ref{fig:tds}(B). The first and most concentrated drops of espresso have a high TDS of ca. 25\% which, after a stable phase of ca. 20 s, rapidly decreases, reaching values close to zero at about 60 s. {Motivated by standard sigmoidal breakthrough-curve fits in packed-bed transport~\cite{Hu2019}}, we propose an empirical relation to describe the data
\begin{equation}\label{eq:tds}
\mathrm{TDS}(t) =  \frac{k_\text{t}}{2} \left[1 - \tanh\left(\frac{t - \ell_\text{t}}{m_\text{t}}\right)\right],
\end{equation}
sketched as a dashed line in Fig.~\ref{fig:tds}(B), with the fitting parameter $k_\text{t}=25.6\pm 1.3$\%, a time offset $\ell_\text{t}=20.9\pm0.9$ s and a decay time of $m_\text{t}=8.8\pm1.5$ s.  Knowing the time-dependent flow rate $Q(t)$ from an experimental run, we can thus obtain the {solute flow rate} ${\dot{m}_\text{d}(t)}$ into the cup by multiplying ${\dot{m}_\text{d}(t) =} \text{TDS}(t)\times Q(t)$, as presented in Fig.~\ref{fig:time_dependent}(C). While the flow rate picks up and increases monotonically during the typical brewing time, the solute flux has a maximum ca. 30 s. The solute flow is thus initially limited by the low flow rate, while at longer times it is limited by the availability of soluble material inside the coffee puck. {We note that while the typical flow rate in the transient dissolution regime is of the order of 1 g/s, the corresponding mass flux of solutes remains in the range of ca. 0.1 g/s, thus supporting the assumption of slow erosion of the puck. The total mass removed in the process amounts to ca. 10\% of the initial mass of the dry coffee puck (18.5 g).} We calculate the total mass of solids removed from the coffee puck, $m_\text{d}(t)$ {by integrating} the solute flux in Fig.~\ref{fig:tds}(D). Again, the sigmoidal shape of this curve can be accurately represented by an empirical relationship
 \begin{equation}\label{eq:mass_flux}
     m_\text{d}(t) =  \frac{k_\text{m}}{2} \left[1 + \tanh\left(\frac{t - \ell_\text{m}}{m_\text{m}}\right)\right],
 \end{equation}
where $k_\text{m}$ is the total soluble mass, and $\ell_\text{m}$ and $m_\text{m}$ represent the time offset and the characteristic time scale of dissolution, respectively. {By fitting the empirical relationship \eqref{eq:mass_flux} we reduce the phenomenological description of the dissolution kinetics to 3 constants, eliminating time dependent measurements from model predictions.} These three parameters are estimated by fitting the relationship~\eqref{eq:mass_flux} to the experimental data, and in this case read $k_\text{m} =2.257\pm0.001$ g, $\ell_\text{m}=19.83\pm0.01$ s, and $m_\text{m}=9.34\pm0.02$ s. We note that, since the underlying temporal dynamics are governed by the dissolution kinetics, the time scales and offset are nearly identical, so one can effectively assume $\ell_\text{m}\approx\ell_\text{t}$ and $m_\text{m}\approx m_\text{t}$. Motivated by these observations, we propose a model to quantify the dissolution process, described in Sec.~\ref{sec:theory}.B. In essence, the dissolution-induced removal of mass can be related to an increase in effective porosity of the medium, which in turn affects its permeability. Assuming the instantaneous porosity $\Phi(t)$ to be proportional to the relative dissolved mass of coffee, Eq.~\eqref{eq:time-dependent-q} provides us with a predictive tool to estimate the time-dependent flow during brewing, where the only parameters are estimated from the long-time equilibrium calibration pressure and flow, $P_c$ and $Q_c$, and the long-time porosity $\Phi_\text{m} = \lim_{t\to\infty} m(t)/m_0 = k_\text{m}/m_0$. For convenience, we repeat the values here: $P_c=12\pm3$ bar, $Q_c=1.90\pm0.15$ g/s; The estimated equilibrium porosity is $\Phi_\text{m}\approx0.122$. The results of such calculations are presented in Fig.~\ref{fig:porosity_model}(A) for a representative example of a typical brewing pressure of 9 bar. Although the minimal model deviates from the flow rate measurement at short times, where swelling and air removal complicate the dynamics, it seems to approximately grasp the flow curve. A similar comparison for other values of brewing pressure (with colour coding as in Fig.~\ref{fig:time_dependent}) in Fig.~\ref{fig:porosity_model}(B) shows that the model fails to grasp the quantitative dynamics for the lowest brewing pressures, when the puck satisfies Darcy's law, but becomes increasingly accurate at higher pressures, providing quantitative predictions of the flow rate dynamics. 

\begin{figure}[t]
\includegraphics{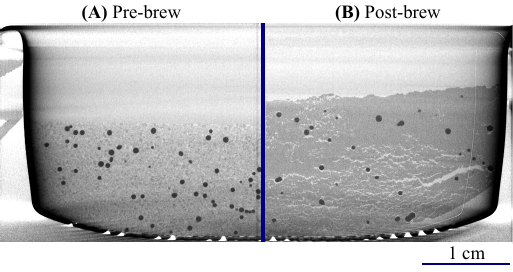} 
    \caption{\textbf{X-ray micro-CT slice through an espresso puck.} 
    \textbf{(A)} Tamped puck before brewing. Glass microbeads (dark) are added for visualisation. \textbf{(B)} Coffee puck after brewing. The post-brew image reflects two distinct structural processes: microscopic particle-scale swelling upon wetting, and macroscopic horizontal delamination and partial lift-off of the bed from the bottom filter mesh, with a layer of air left in between.} 
    \label{fig:tomography}
\end{figure}

Although the time scales of the brewing process and the structure of a typical espresso machine preclude direct visualisation of flow in situ, we performed X-ray $\mu$-CT scans of a coffee puck after sample preparation and shortly after brewing. An overlay of typical images is presented in Fig.~\ref{fig:tomography}. The pre-brew puck in panel (A) reveals a tightly packed structure with evident small-scale heterogeneity. A comparison of the sample thickness with the same puck imaged after brewing in panel (B) shows a considerable swelling but also a reconfiguration of the sample. In the post-brew tomographic slice, multiple horizontal delaminations and cracks are seen. In the bottom part, close to the filter mesh, the coffee puck has peeled off the basket and a void space was created in between the filter mesh and the puck. However, we associate this with the operation of the overpressure valve which may reverse the direction of flow for an instant after the brewer flow is turned off to relieve the built up pressure. This short backflow might lead to a reconfiguration of the sample after brewing has finished. 

\begin{figure}
    \includegraphics{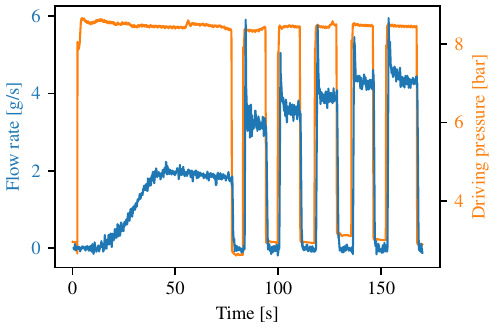}   
    \caption{
    \textbf{Flow rate changes upon repeated brewing.}
    Restarting the brew cycle several times increases the flow rate without causing additional dissolution. This effect arises from the rapid reorganization of the coffee puck induced by partial backflow through the three-way valve found in all modern espresso machines, which evacuates water from the puck’s surface after brewing. The rapid flow changes produce the delaminations seen in tomographic images in Fig.~\ref{fig:tomography}. } 
    \label{fig:brew_restarting}
\end{figure}

To explore the effect of the delaminations on the brewing process, we additionally performed an experiment where the coffee puck was first exposed to a long brewing process when pressure of 8 bar was applied for ca. 75 s, followed by repeated cycles of turning the brewer on and off. The pressure signal, together with the resulting flow rate is plotted in Fig.~\ref{fig:brew_restarting}. Upon repeated brewing, we witness a significant increase of the flow rate without additional dissolution, which suggests that a reconfiguration of the sample takes place each time the pressure is applied. The driving pressure, indicated at the pump, switches between the working pressure of ca. 8 bar and no flow (closed valves) at ca. 3 bar. Upon switching off the pump, there is a slight backflow due to the presence of a three-way valve in the hydraulic circuit above the coffee puck. We argue that the delaminations that occur when the brewing process stops contribute to an increasing channelling in the sample, a process long known in the coffee community to negatively affect the quality of the brew. In a channelled coffee, the flow through a puck is focused in one or several channels instead of uniformly penetrating the sample. As a result, even at relatively high flow rates, the extraction yield of coffee remains low. The evidence in Fig.~\ref{fig:brew_restarting}, although presented with no visualisation of internal reconfiguration, suggests that any brewing process that involves a temporary pause or repeated application of pressure might be prone to channelling and therefore should be avoided. 

\section{Discussion}\label{sec:discussion}

{For convenience, we summarise the parameters of the description, measured quantities, material characteristics and fitted parameters in Table~\ref{tab:parameters} in Appendix \ref{sec:appendix}. Below, we discuss in details the limitations of the presented approach and emphasise the complexity of espresso brewing as a multiphysics phenomenon.}

\subsection{Ground coffee as a poroelastic medium}

When coffee is treated as a porous medium, a crude estimation of the flow rate directly from the Darcy' law, Eq.~\eqref{eqn:darcy-classical}, with the Carman-Kozeny closure, Eq.~\eqref{eqn:Darcy_Carman_Kozeny}, taking $d_\mathrm{p} = 200\ \micron$ (the typical grind size for coffee) yields a flow rate of the order of 2000 g/s at $\Phi = 0.5$, thus overestimating the typical espresso flow rate by a factor of $10^3$. This is consistent with our control experiments, where we packed the coffee basket with an equivalent volume of $200\ \micron$ diameter glass beads, which at typical brewing pressures give virtually no resistance to flow compared to coffee grounds of similar size. Moreover, the flow rate through the glass is weakly dependent on the water temperature. On the other hand, we note that the Carman-Kozeny relationship is of little predictive value in typical applications, and we expect it to break down completely when the medium is close to clogging and $\Phi\to0$, which is a limitation of the presented simple model \cite{Tien2013}. The universal character of the behaviour of permeability at low porosities could be the result of the chosen constitutive law. A possible improvement would be to consider different models for the behaviour of permeability near the percolation threshold. {Continuum-scale, volume-averaged descriptions of coffee extraction have also been employed in the food engineering literature. For example, Sano {\it et al.}~\cite{sano_2019} develop a porous-medium extraction model using effective transport coefficients, demonstrating that macroscopic flow and dissolution trends in packed coffee beds can be captured despite limited pore-scale scale separation. In this spirit, the permeability $k$ in our model should be interpreted as an effective hydraulic property inferred from long-time measurements, rather than a pore-scale constitutive constant.}

{In a more general context, stress-dependent constitutive relations for permeability and stiffness are well established in soil mechanics and poromechanics, where compaction leads to non-linear evolution of porosity, tortuosity, and hydraulic conductivity under effective stress~\cite{Mitchell2025,Coussy2003}. Such frameworks often employ empirically calibrated stress--permeability laws beyond the Carman--Kozeny form~\cite{terzaghi1943theoretical,Wang2000}. Extending the present model in this direction would be a natural next step, but requires independent constitutive measurements for wetted coffee grounds, whose evolving microstructure is additionally influenced by swelling and dissolution.}

During brewing, coffee undergoes significant changes; Spent coffee pucks resemble sugar cubes or dry cookies that can be carefully handled but crumble under moderate stress. This behaviour contrasts strongly with that of dry coffee grounds. The source of this difference is the significant swelling that occurs when coffee is in contact with hot water (cf. Fig. \ref{fig:tomography}), which in conjunction with mechanical confinement leads to a large decrease of the effective pore diameter. A 1000-fold decrease in permeability can be explained by a 30-fold decrease in effective pore diameter, which leads to estimated pore sizes of the order $6\  \micron$. Such crude estimates correspond well to microporosity observations performed using electron tomography \cite{Lindsey_2024}.

At the same time, we argue in this contribution that elasticity of the coffee puck is an important factor that determines the flow response. While the proposed {phenomenological} model circumvents the need for the determination of Young's modulus for coffee, instead fitting the long-time calibration pressure and flow, an independent measurement of $Y$ would {be} useful for independent validation. However, to the best of our knowledge, no such mechanical measurements have been conducted on wetted ground coffee. As a reference point, measurements of elastic properties of roasted coffee beans of Severin et al. \cite{Severin_2024} report forces of around $50$ N at a displacement of $0.15$ mm. Taking the size of a coffee bean to be $1 \mathrm{cm}$ and an effective cross section $80\ \mathrm{mm}^2$ yields Young's modulus of $40 \mathrm{MPa} = 400 \mathrm{bar}$ for whole roasted beans. 

{An important assumption in our model is that the thermophysical properties of water remain approximately constant during espresso extraction. The dynamic viscosity of liquid water over the narrow brewing temperature window of $90-96^\circ\text{C}$, varies on the order of ~5~\%~\cite{CRC2016}. Such moderate changes primarily rescale the overall hydraulic resistance and can be effectively absorbed into the calibrated reference flow scale, without affecting the qualitative non-linear pressure–flow saturation behaviour discussed here. Moreover, in the discussed range of temperatures and pressures, one can assume water to be incompressible. A more comprehensive multi-physics model could incorporate temperature-dependent constitutive relations, but this would require independently measured material responses for wetted coffee beds, for which the available data are scarce.} {While said measurements remain challenging, we stress, however, that the Young's modulus entering the model in Sec. \ref{sec:theory} should be understood as an effective parameter at the scale of the coffee puck. Its role in the present model is to set the characteristic pressure scale for compaction. The emergence of flow saturation and the shape of the long-time pressure-flow relation are controlled primarily by the ratio of the applied pressure to this scale, $P/Y$, and are therefore robust to moderate uncertainty in the precise value of $Y$.}

\subsection{Solubility kinetics of ground coffee}

The model proposed in this work, coupled with experimental results, shows that the time-dependent kinetics of the coffee dissolution are coupled to the flow dynamics through this reactive medium, rendering a complex response in terms of the mass flux of solutes. The results of our TDS measurements in Fig.~\ref{fig:time_dependent} support the claim that during a typical espresso brewing the majority of readily extractable compounds are dissolved almost instantaneously \cite{Zanoni1992,Spiro1993}, and are subsequently removed by the increasing flow. {Because the flow rate is not strictly constant during extraction, one may also analyse solute release as a function of cumulative extracted volume rather than time. We have verified that this alternative representation yields the same qualitative trends.}

We introduced the time-dependent porosity, $\Phi(t)$, into the flow model to obtain predictions for the time dependence of the flow rate $Q$ at different driving pressures (cf. Fig.~\ref{fig:time_dependent}), which are in qualitative agreement for both high and low pressures, reproducing the initial delay between wetting and onset of the flow for higher pressures. Our simplistic model predicts a complete shut-off of the flow at $P_{\mathrm{ext}} / Y > \Phi$. This might be relevant for the initial stage of brewing, where the value of $\Phi$ is low, and would provide a qualitative explanation of the delayed onset of flow, but we note that this phase is complicated by the multi-phase character of flow and swelling.  {A more direct comparison between the experimentally measured dissolution dynamics and the model provides additional insight into the transient evolution of the flow. The time-resolved flow measurements in Fig.~\ref{fig:time_dependent} exhibit a rapid initial increase on a characteristic timescale of $10$--$20$~s, followed by a slower evolution towards a plateau. Importantly, this timescale is comparable across different applied pressures, despite significant differences in the long-time flow rate. This observation suggests that the kinetics of solute release are primarily governed by the properties of the coffee matrix rather than by the instantaneous flow conditions. Within our framework, dissolution enters through the experimentally inferred dissolved mass fraction $\Phi(t)$, which modulates the effective permeability. In this way, the model captures the coupling between chemical mass removal and hydraulic response: dissolution gradually opens the pore space, while poroelastic compaction acts in the opposite direction. The resulting competition leads to the observed time dependence of the flow rate. We emphasize that the hyperbolic tangent form used to represent $\Phi(t)$ is not intended as a mechanistic kinetic law, but rather as a compact phenomenological representation of the available soluble fraction, sufficient to reproduce the experimentally observed trends.} Beyond a qualitative model which describes the initial slow percolation, the detailed modelling of this phase is beyond the scope of this manuscript and requires higher time resolution of flow, TDS and pressure data.

{The general coupling between dissolution, evolving porosity, and permeability has been studied extensively in the reactive transport literature, for example in carbonate dissolution during CO$_2$ injection and matrix acidizing. Continuum-scale reactive transport models integrate advective--diffusive transport with interfacial reaction kinetics to describe how removal of solid mass alters pore structure and hydraulic properties over time \cite{Steefel2005,Ladd_2021}. While such approaches are often formulated for rigid porous matrices, they provide a useful conceptual framework for interpreting the dissolution-driven evolution of permeability in espresso extraction, where chemical opening competes with simultaneous poroelastic compaction of the deformable granular bed. In this sense, our closure through $\Phi(t)$ represents a coarse-grained reactive transport description, modified by the additional constraint of stress-dependent compaction in the coffee bed.}

The dissolution process is crucially dependent on the structure of the porous medium. In the case of coffee, the bi-modal distribution of grain sizes (cf. Fig.~\ref{fig:abundance_vs_size}) is universal across different grind settings and devices \cite{Gagne_2025}. Recently, an approach with sieving grounds to achieve a unimodal (single-peak) distribution has been proposed \cite{Lindsey_2024}. What is interesting, Smrke et al. \cite{Smrke_2024} find that, for a given coffee, the extraction time of an espresso can be predicted by knowing the share of fines and the size of the large particles (boulders).
The effect of increased surface area due to a smaller particle size for coffee with a high fraction of fines on the extraction efficiency seems to be marginal.

Finally, we note that in Refs. \cite[ch. 7.5.7]{Illy_2004} and
\cite{Fasano_2000, Hoffmann_2018} the authors state that the flow during espresso brewing does not linearly depend on pressure for pressures around 9 bar, and that an increase in pressure actually causes a decrease in the average flow, despite Darcy's law suggesting otherwise -- we corroborate these findings and propose an explanation of the mechanism underlying this behaviour. However, other mechanisms have been proposed to explain the complex behaviour of coffee, including the migration of coffee fines within the basket, jamming, or the formation of preferential flow paths that change the effective pore sizes in the medium. Future experiments, in particular time-resolved measurements of structural changes in the coffee puck might bring new evidence to examine these hypotheses. 

{\subsection{Ground coffee as a swelling medium}}

{Beyond poroelastic compaction, the permeability of the coffee bed is strongly influenced by swelling of the solid matrix upon wetting. Roasted coffee particles absorb water rapidly, leading to hygroscopic expansion of the polysaccharide-rich cellular structure. Under mechanical confinement in the portafilter basket, this swelling reduces the available void space and can substantially decrease the effective pore diameter and permeability. Particle size distribution analysis by Hargarten {\it et al.}~\cite{Hargarten_2020} yielded a total increase of the particle diameter by approximately $15\%$. This value is close to the results obtained by Spiro {\it et al.}~\cite{Spiro1993}, but higher than subsequent laser-diffraction measurements reported by other authors.}

{Swelling does not act in isolation, but competes with other mechanisms that reshape the porous matrix during extraction. In particular, recent pore-scale simulations by Matias {\it et al.}~\cite{Matias_2021} emphasize the interplay between swelling and erosion in determining the steady-state structure of the bed under an imposed pressure drop. In their framework, swelling reduces local shear stresses by constricting the pore space, whereas erosion acts in the opposite direction by reopening flow pathways and increasing shear. The resulting dynamics depend sensitively on the erosion threshold: low thresholds lead to erosion-dominated regimes, while high thresholds favour swelling-controlled compaction, with a sharp transition between the two behaviours. This highlights that small changes in the balance of swelling and erosion can strongly affect the long-time porosity and permeability of the coffee bed.}

{Swelling-induced pore constriction bears qualitative similarities to hydration effects in clay-rich soils, where permeability reductions under confinement are well documented. In clays, swelling is often driven by osmotic interlayer hydration and may involve structural dissociation~\cite{norrish1954swelling}. In contrast, swelling in roasted coffee grounds reflects a finite relaxation of a cross-linked biopolymer matrix upon water contact~\cite{Hargarten_2020}. Importantly, this process occurs rapidly on the timescale of seconds after wetting, and therefore contributes dynamically during brewing rather than being solely a post-extraction phenomenon.}

{The micro-CT images in Fig.~\ref{fig:tomography} highlight this evolving structure. The post-brew scans reflect both microscopic particle-scale swelling and macroscopic delamination or partial lift-off of the puck. The latter is likely promoted by rapid depressurization and transient backflow at the end of the shot, whereas swelling itself occurs immediately upon infiltration. We ascribe the observed horizontal cracks and peeling features as signatures of post-shot stress release, where confinement by the side walls primarily permits expansion in the vertical direction.}

{These observations reinforce that espresso extraction involves an evolving porous structure, in which swelling, confinement, and pressure-driven deformation jointly regulate permeability. While the present model adopts a coarse-grained description of this evolution, the results of Matias {\it et al.}~\cite{Matias_2021} and the microstructural evidence presented here suggest that future work incorporating coupled swelling--erosion dynamics and stress-dependent permeability laws could further improve predictive descriptions of extraction dynamics.}

\section{Conclusions}\label{sec:conclusions}

The flow of hot water under pressure through a tightly packed bed of ground coffee is a phenomenon of paramount importance for many amateurs of the beverage {around the world. For} a scientist, coffee is a beverage of many faces, from the cultural perspective, through the chemical complexity and biological effect, to the solid and fluid mechanics of the preparation process. From the physics point of view, it is a fascinating multi-physics and multi-scale system, which can be approached using a variety of theoretical and experimental tools.

Motivated by the non-linear flow response of coffee to the driving pressure and scarce evidence, in this study we perform systematic experiments to explore this dependence using a café-grade espresso machine, augmented with pressure and mass sensors. The examination of flow rate at long brewing times allows us to observe the saturated, equilibrium behaviour of flow rate after the dissolution process of coffee solubles is complete. Our measurements reveal a clear deviation from the classical Darcy behaviour, particularly at higher pressures, where the flow rate does not scale linearly with pressure. To rationalise these observations, we propose a minimal {coarse-grained phenomenological description} that assumes poroelasticity of the coffee bed, and captures the essential features of the pressure-dependent flow rate. Further, by measuring time-resolved solute concentrations, we show that the temporal evolution of the flow is governed primarily by the dissolution dynamics rather than a structural rearrangement of the porous matrix. These insights establish that espresso brewing is a self-regulating process where solute dissolution and matrix elasticity interact to shape the overall extraction dynamics, and suggest that future optimisation efforts should consider the coupled physico-chemical nature of the coffee bed, and that the assumption of constant bed permeability should be relaxed to formulate realistic predictions.

{Finally, we emphasise the limitations of the present coarse-grained framework. 
First, the model focuses on the long-time, saturated flow regime and does not resolve the early-stage multiphase dynamics associated with air displacement, wetting, and swelling. 
Second, permeability is described using an effective constitutive relation (Carman--Kozeny), and spatial heterogeneity, fines migration, and channel formation are not explicitly treated. 
Third, the mechanical response of the coffee bed is represented by an effective poroelastic modulus, rather than a directly measured, time-dependent material property of the evolving granular medium. 
Fourth, dissolution is incorporated through a phenomenological, experimentally inferred porosity evolution $\Phi(t)$, rather than a first-principles reactive transport model that couples chemistry, transport, and mechanics at the pore scale. 
In addition, the chemical complexity of coffee, involving many soluble species with distinct kinetics, is reduced here to a single effective scalar quantity (TDS).} 

{These simplifications allow us to isolate and quantify the role of poroelastic compaction in regulating the macroscopic pressure--flow response, but a more complete description of espresso extraction would require a fully coupled multiphysics treatment combining two-phase flow, reactive transport, and evolving microstructure.}

{
The experimental results presented here correspond to brewing conditions typical of café practice, including grind size, roast level, dose, and applied pressure. A systematic exploration of how variations in these parameters affect the flow response and extraction dynamics remains an interesting direction for future work.
}

\section{Acknowledgments}
ML, RW and PS acknowledge funding from the University of Warsaw IDUB programme. The work of FM was partially supported by a DAAD scholarship for master's studies in Germany. We thank Dorota Nowicka and Kamil Kolasa for their contributions to the early versions of the experiments. The project benefitted from guidance, support, and equipment from coffee industry specialists, including in particular (in alphabetical order): Krzysztof Blinkiewicz, Łukasz Jura, Bartosz Krzymowski, Rafael Młodzianowski, Kacper Ornat, Justyna Sarnowska, Wojtek Szymański, Maciej Zielenkiewicz.

We are indebted to Coffeelab Warsaw (\url{coffeelab.pl}) for providing high quality coffee beans, barista and sensory training, as well as technical support. Coffee Machines Sale (\url{cmsale.com}) is gratefully acknowledged for access to barista-grade equipment essential for reproducible espresso brewing, measurements, and invaluable know-how.

\section*{Data Availability}
All data and analysis codes used in this study are publicly available. The source code and datasets are hosted on GitHub \cite{Waszkiewicz_2025_github}, with an archived, citable version available on Zenodo \cite{Waszkiewicz_2025_zenodo}.

\section*{Author Contributions}

\noindent
\textbf{Conceptualization}: RW, ML (lead), FM, MD, PS (supporting).
\textbf{Data curation}: RW, ŁB, MPM, ML (supporting).
\textbf{Formal analysis}: RW (lead), FM, ŁB, MPM, PS, ML (equal).
\textbf{Funding acquisition}: RW, ML (lead).
\textbf{Investigation}: RW, ML (equal), FM, ŁB, MPM, MD (supporting).
\textbf{Methodology}: RW, ML (lead), MD, PS (supporting).
\textbf{Project administration}: ML (lead), RW (supporting).
\textbf{Resources}: ML (lead), FM (supporting).
\textbf{Software}: ŁB (lead), RW (equal).
\textbf{Supervision}: RW, ML (lead).
\textbf{Validation}: ML (supporting).
\textbf{Visualization}: RW (lead), FM, ŁB, MPM (supporting).
\textbf{Writing -- original draft}: RW, ML (lead), FM (supporting).
\textbf{Writing -- review \& editing}: RW, ML (lead), FM, ŁB, MPM, MD, PS (supporting).

\onecolumngrid
\newpage
\appendix 

{\section{Summary of parameters} \label{sec:appendix}
\centering
\captionof{table}{Summary of symbols and parameters}
\begin{tabular}{lccl}
\toprule
Parameter & Symbol  & Type & Value \\
\midrule\midrule
basket area & $A$ & m &  $2.64\times10^3$ mm$^2$ (diameter 58 mm) \\
initial bed thickness & $h_0$ & m & 14 mm \\
initial mass & $m_0$ & m & $18.50 \pm 0.05$ g \\
applied pressure & $P$ & m & $1$ bar -- $12$ bar \\
total dissolved solids & TDS & m & $<25\%$ \\ \midrule
fluid viscosity & $\mu$ & c &  \\
Young's modulus & $Y$ & c & \\
stress-free porosity & $\Phi$ & c & \\
pore diameter & $d_p$ & c &  \\
reference flow rate & $Q_{\text{ref}}$ & c & $YAd_p^2/(\mu h_0)$ \\ \midrule
vertical coordinate & $z$ &v& \\ 
superficial velocity & $u(z)$ & v & \\
local pressure & $p(z)$ & v & \\
pressure difference & $\Delta P$ & v & \\
matrix stress & $\sigma(z) $ & v & \\
bed porosity & $\phi(\sigma)$ & v & \\
permeability & $k(\phi)$ & v & \\
strain & $e(z)$ & v & \\
dissolved mass & $m_\text{d}(t)$ & v & \\
maximum flow rate & $Q_\text{m}(\Phi)$ & v & \\
instantaneous stress-free porosity & $\Phi(t)$ & v & \\
flow rate & $Q(t)$ & v & $\sim 1$ g/s \\
solute flux & $\dot{m}_\text{d}$ & v & $\sim 0.1$ g/s \\
dimensionless pressure & $P^*$ & v & $P/Y$ \\
normalised pressure & $\hat{P}$ & v & $P/(Y\Phi)$ \\
normalised flow rate & $\hat{Q}$ & v & $Q/Q_\text{m}$ \\
 \midrule
calibration flow rate & $Q_c$ & f & $1.90 \pm 0.15$ g/s \\
calibration pressure & $P_c$ & f & $12 \pm 3$ bar \\
TDS amplitude & $k_\text{t}$ & f & $25.6 \pm 1.3$\% \\
TDS time offset & $\ell_\text{t}$ & f & $20.9 \pm 0.9$ s \\
TDS decay timescale & $m_\text{t}$ & f & $8.8 \pm 1.5$ s \\
total dissolved mass & $k_\text{m}$ & f & $2.257 \pm 0.001$ g \\
mass time offset & $\ell_\text{m}$ & f & $19.83 \pm 0.01$ s \\
mass timescale & $m_\text{m}$ & f & $9.34 \pm 0.02$ s \\
long-time stress-free porosity & $\Phi_\text{m}$ & d & $  k_\text{m}/m_0 \approx 0.122$  \\
\bottomrule
\end{tabular}
\label{tab:parameters}
\begin{flushleft}
Notes: m -- measured; c -- constant; v -- variable; f -- fitted; d -- derived from fit;
\end{flushleft}}
\twocolumngrid

\bibliography{sources}

\end{document}